\documentclass[11pt,a4paper]{article}
\usepackage{graphicx}
\usepackage{jcappub}

\def\sun{\hbox{$\odot$}}

\def\la{\mathrel{\mathchoice {\vcenter{\offinterlineskip\halign{\hfil
$\displaystyle##$\hfil\cr<\cr\sim\cr}}}
{\vcenter{\offinterlineskip\halign{\hfil$\textstyle##$\hfil\cr
<\cr\sim\cr}}}
{\vcenter{\offinterlineskip\halign{\hfil$\scriptstyle##$\hfil\cr
<\cr\sim\cr}}}
{\vcenter{\offinterlineskip\halign{\hfil$\scriptscriptstyle##$\hfil\cr
<\cr\sim\cr}}}}}
\def\ga{\mathrel{\mathchoice {\vcenter{\offinterlineskip\halign{\hfil
$\displaystyle##$\hfil\cr>\cr\sim\cr}}}
{\vcenter{\offinterlineskip\halign{\hfil$\textstyle##$\hfil\cr
>\cr\sim\cr}}}
{\vcenter{\offinterlineskip\halign{\hfil$\scriptstyle##$\hfil\cr
>\cr\sim\cr}}}
{\vcenter{\offinterlineskip\halign{\hfil$\scriptscriptstyle##$\hfil\cr
>\cr\sim\cr}}}}}

\title{An asteroseismic constraint  on the mass of the  axion from the
  period drift of the pulsating DA white dwarf star L19$-$2}
  
\author[a,b]{Alejandro H. C\'orsico,}
\author[c]{Alejandra D. Romero,}
\author[a,b]{Leandro G. Althaus,}
\author[d,e]{Enrique Garc\'\i a--Berro,}
\author[e,f]{Jordi Isern,} 
\author[c]{S. O. Kepler,}
\author[a,b]{Marcelo M. Miller Bertolami,}
\author[g]{Denis J. Sullivan,}
\author[g,h]{Paul Chote}
 
\affiliation[a]{Grupo de Evoluci\'on Estelar y Pulsaciones,
                Facultad de Ciencias Astron\'omicas y Geof\'{\i}sicas,  
                Universidad  Nacional de La Plata,
                Paseo del  Bosque s/n,  
               (1900) La Plata, 
                Argentina}
\affiliation[b]{Instituto de Astrof\'{\i}sica La Plata, 
                CONICET-UNLP, Argentina}
\affiliation[c]{Departamento de Astronomia,
                Universidade Federal do Rio Grande do Sul,
                Av. Bento Goncalves 9500, Porto Alegre 91501-970, RS, 
                Brazil}
\affiliation[d]{Departament de F\'\i sica Aplicada, 
                Universitat Polit\`ecnica de Catalunya, 
                c/Esteve Terrades, 5,  
                08860 Castelldefels,  
                Spain}
\affiliation[e]{Institute for Space Studies of Catalonia, IEEC,
                c/Gran Capit\`a 2--4, Edif. Nexus 104,   
                08034  Barcelona, 
	        Spain}
\affiliation[f]{Institut de Ci\`encies de l'Espai, CSIC, 
                Campus UAB, Facultat de Ci\`encies, Torre C-5, 
                08193 Bellaterra, 
                Spain}
\affiliation[g]{School of Chemical \& Physical Sciences, 
                Victoria University of Wellington,
                Wellington, New Zealand}
\affiliation[h]{Department of Physics, University of Warwick,
                Coventry CV4 7AL, United Kingdom}

\emailAdd{acorsico@fcaglp.unlp.edu.ar,
          alejandra.romero@ufrgs.br, 
          althaus@fcaglp.unlp.edu.ar,
          enrique.garcia-berro@upc.edu,
          isern@ieec.cat,
          kepler@if.ufrgs.br,
          mmiller@fcaglp.unlp.edu.ar,
          denis.sullivan@vuw.ac.nz, 
          p.chote@warwick.ac.uk}

\abstract{We  employ  an  asteroseismic   model  of  L19$-$2,  a
  relatively massive ($M_{\star} \sim 0.75 M_{\sun}$) and hot ($T_{\rm
  eff} \sim 12\,100$  K) pulsating DA (H-rich  atmosphere) white dwarf
  star (DAV  or ZZ Ceti variable),  and use the observed  values of the
  temporal rates of period change of its dominant pulsation modes ($\Pi
  \sim 113$~s and $\Pi \sim 192$~s), to derive a new constraint on the
  mass of the axion, the hypothetical non-barionic particle considered
  as a possible component of the  dark matter of the Universe.  If the
  asteroseismic model employed is  an accurate representation of
  L19$-$2, then  our results indicate
  hints of extra cooling in this star, compatible with emission
  of axions of mass $m_{\rm  a} \cos^2 \beta \lesssim 25$~meV
  or an axion-electron coupling constant of
   $g_{\rm ae} \lesssim 7 \times 10^{-13}$.}

\notoc

\keywords{elementary  particles  --   stars:  oscillations  --  stars:
  individual: ZZ Ceti stars -- stars: white dwarfs}
  
\begin{document}

\maketitle  

   
\section{Introduction and context}  
\label{intro}  

Charge-parity  (CP) symmetry  \cite{L54,P54} states  that the  laws of
physics  should be  the same  if  particles were  switched with  their
antiparticles (C symmetry) and their spatial coordinates were
inverted  (P symmetry).   The theory  of quantum  chromodynamics (QCD)
predicts   that   the  CP   symmetry   must   be  broken   in   strong
interactions. However,  this violation is  not observed in  nature, at
variance   with    what   happens    in   the    electro-weak   theory
\cite{1973PThPh..49..652K}.  This constitutes  a long-standing problem
in   particle   physics,   known   as  the   ``strong   CP''   problem
\cite{2010RvMP...82..557K}.

A possible solution  to strong CP problem was proposed  about 40 years
ago  by  postulating the  existence  of  a  new particle,  the  pseudo
Nambu-Goldstone  boson  ---  that  is,  a  boson  with  a  small  mass
\cite{1996slfp.book.....R}       ---        known       as  an   axion
\cite{1977PhRvL..38.1440P,1978PhRvL..40..223W,1978PhRvL..40..279W}.
At present, however, this particle  has not been detected.  Currently,
axions are  the target  of many  theoretical and  experimental studies
aimed at  proving their existence.  One the main interests  driving these
experiments  is  that axions  are  candidates  for  being one  of  the
possible components  of cold dark  matter (CDM).  The  contribution of
axions     to    CDM,     however,    depends     on    their     mass
\cite{2007JPhA...40.6607R},  that is  a free  parameter of  the theory
that postulates  their existence.   Axions are  neutral (they  have no
electric charge), and they interact very weakly with normal matter and
radiation. Among the various types  of axion models, the most commonly
discussed    in     the    literature     are    the     KVSZ    model
\cite{1979PhRvL..43..103K,1980NuPhB.166..493S},   where   the   axions
couple   with    photons   and    hadrons,   and   the    DFSZ   model
\cite{1981PhLB..104..199D,Z80},  where  they  also couple  to  charged
leptons, like electrons.  In this work,  we focus on DFSZ axions, that
is those that interact with electrons.

The coupling strength of DFSZ axions to electrons is defined through a
dimensionless coupling constant, $g_{\rm ae}$, which is related to the
mass   of    the   axion,   $m_{\rm   a}$,    through   the   equation
\cite{1996slfp.book.....R}:

\begin{equation}
g_{\rm ae}= 2.8 \times 10^{-14}\  \frac{m_{\rm a} \cos^2 \beta}{1\ {\rm meV}},
\label{eq1}
\end{equation}

\noindent where  $\cos^2 \beta$ is a  free, model-dependent, parameter
that is usually  set equal to unity.  Since the  theory does not place
any constraint on the mass of  axions, it must be inferred from either
sophisticated  terrestrial experiments  \cite{2016arXiv160100578R}, or
indirectly    by    using    well-studied    properties    of    stars
\cite{1996slfp.book.....R}.  In  this paper, we use  white dwarf stars
to constrain the mass of the axion.

White dwarfs  are the very  late evolutionary  state of most  low- and
intermediate-mass stars  ($M_{\star} \sim 0.8-10  M_{\sun}$), including
our Sun \cite{review}.  Because in the degenerate core of white dwarfs
free electrons  are abundant,  axions would  be copiously  produced in
their  interiors  \cite{R86}.   Moreover,  because  white  dwarfs  are
strongly degenerate and  do not have relevant  nuclear energy sources,
their  evolution consists  in  a  slow cooling  process  in which  the
gravothermal energy  release is the  main energy source  driving their
evolution  \cite{1952MNRAS.112..583M}.  Thus,  any additional  cooling
mechanism should  be prominent, and have  observable consequences.  At
the typical temperatures and densities ($\log T \sim 6, \log \rho \sim
6$) found  in the cores of  white dwarfs, the emission  of DFSZ axions
would  take  place in  the  deepest  regions  of these  stars  through
Compton, pair annihilation, and  bremmsstrahlung processes, this
last mechanism being the dominant one \cite{R86}.   Under these
conditions, the axion emission rate is given by \cite{Nea87,Nea88}:

\begin{equation}
\epsilon_{\rm a}= 1.08 \times 10^{23} \frac{g^2_{\rm ae}}{4\pi} 
\frac{Z^2}{A} T_7^4 F(T, \rho)\ \ [{\rm erg/g/s}]. 
\label{eq2}
\end{equation}

\noindent  where  $Z$  and  $A$  are  the  atomic  and  mass  numbers,
respectively, $T_7$ is  the temperature in units of $10^7$  K, and $F=
F(T, \rho)$  is a function of  the density and temperature  that takes
into account  the Coulomb plasma  effects. As  can be seen,  the axion
emission is determined by the coupling strength. In turn, the strength
of the coupling of axions with  electrons depends on the axion mass --
see Eq.~(\ref{eq1}).   Consequently, the more massive  the axions are,
the larger  the axion emission  is.  Since axions can   freely
escape  from  the  interior  of white  dwarfs,  their  emission  would
accelerate cooling, with more  massive axions producing larger cooling
rates.

One  of the  possible  methods  to determine  the  mass  of the  axion
consists in using the white  dwarf luminosity function. The luminosity
function of white dwarfs is defined  as the number of white dwarfs per
unit  bolometric magnitude  and unit  volume. Clearly the cooling
rate of white dwarfs will have an impact on the form of the
luminosity function. It was
found \cite{WDLF} some time ago that when axion emissivity is included
in  the white  dwarf evolutionary  models, the  agreement between  the
theoretical  calculations  and  the observed  white  dwarf  luminosity
function   significantly   improves   \cite{WDLF,2009JPhCS.172a2005I}.
These  early computations were enhanced later  on by employing new
white dwarf luminosity functions of the Galactic Disk derived
from different catalogues, and adopting  new  theoretical
cooling sequences constructed on the  basis of full evolutionary white
dwarfs models that incorporate self-consistently the effects of axions
on the thermal  structure of the white dwarf \cite{2014JCAP...10..069M}.
This later study found the need for extra cooling emission from
white dwarfs to be marginal and dependent on the derivation of
the white dwarf luminosity function.

The second  method of interest used  to measure the mass  of the axion
involves using  the rate of  period change of pulsating  white dwarfs.
Pulsating white dwarfs  with H-rich envelopes, also  called ZZ~Ceti or
DAV variable stars,  constitute the most numerous  class of degenerate
pulsators,    with     over    160     members    known     to    date
\cite{2013MNRAS.430...50C}.    ZZ~Ceti  stars   exhibit  multiperiodic
luminosity  changes resulting  from  spheroidal  modes ---  spheroidal
modes   satisfy   $(\vec{\nabla}   \times  \vec{\xi})_r=   0$,   where
$\vec{\xi}$     is     the      Lagrangian     displacement     vector
\cite{1989nos..book.....U}.  The  modes  are  non-radial  gravity-mode
($g$-mode)  pulsations of  low  degree ($\ell  \leq  2$) with  periods
ranging  from 70  to  1500~s  \cite{WK08,FB08,review}.  The  pulsation
periods ($\Pi$)  of ZZ~Ceti  stars experience a  secular drift  as the
stars  cool,   allowing  for   measurable  rates  of   period  change,
$\dot{\Pi}\equiv d\Pi/dt$. Indeed, as the temperature in the core of a
white dwarf decreases,  the plasma increases its  degree of degeneracy
so  the  Brunt-V\"ais\"al\"a  (buoyancy) frequency  ---  the  critical
frequency  of   $g$-mode  pulsations   \cite{1989nos..book.....U}  ---
decreases,  and the  pulsational spectrum  of the  star is  shifted to
longer periods.  On the other hand, residual gravitational contraction
(if  present) acts  in  the opposite  direction,  thus shortening  the
pulsation periods.  The competition  between the increasing degeneracy
and gravitational contraction gives  rise to a detectable $\dot{\Pi}$.
In particular,  it has been  shown \cite{Wingetet83} that the  rate of
change of the  pulsation periods is related to the  rates of change of
the temperature at the region  of the period formation, $\dot{T}$, and
of the  stellar radius, $\dot{R_{\star}}$, according  to the following
order-of-magnitude expression:

\begin{equation}
  \frac{\dot{\Pi}}{\Pi} \approx -a \frac{\dot{T}}{T} +
  b \frac{\dot{R_{\star}}}{R_{\star}},
\label{eq-dotp}
\end{equation}

\noindent where $a$  and $b$ are constants whose values  depend on the
details of the white dwarf modeling  (however, both $a$ and $b \approx
1$).  The first term in Eq.~(\ref{eq-dotp}) corresponds to the rate of
change in period induced by the  cooling of the white dwarf, and since
$\dot{T}<0$, it is a positive contribution. The second term represents
the rate of change due to gravitational contraction ($\dot{R}<0$), and
it is a negative contribution. In principle, the rate of change of the
period can  be measured by  observing a  pulsating white dwarf  over a
long time interval when one or  more very stable pulsation periods are
present  in their  light curves.  In the  case of  pulsating DA  white
dwarfs, cooling  dominates over  gravitational contraction, in  such a
way that the second term in Eq.~(\ref{eq-dotp}) is usually negligible,
and only positive values of the  observed rate of change of period are
expected.

The possibility of employing the measured rate of period change in the
ZZ~Ceti star G117$-$B15A to derive a  constraint on the mass of axions
was raised for the first time by \cite{Primero}.  G117$-$B15A ($T_{\rm
eff}=       11\,430-12\,500$~K,       $\log       g=       7.72-8.03$,
\cite{1995ApJ...438..908R,2000BaltA...9..119K,2001ASPC..226..299K,
1995ApJ...449..258B,2004ApJ...600..404B}) is the  best studied star of
the ZZ  Ceti class, with  pulsation periods at 215.20~s,  270.46~s and
304.05~s  \cite{Kea82}.  The  evolution of  DA white  dwarfs with  and
without  axion  emission   was considered \cite{Primero},
and  the theoretical values of  $\dot{\Pi}$ associated
with the  largest-amplitude  mode  with  period  at  $\sim  215$~s  for
increasing masses of  the axion were compared to the  observed rate of
change of period with time of G117$-$B15A. Employing a
semi-analytical treatment, $m_{\rm   a}   \cos    \beta=   8.7$~meV
($g_{\rm ae}= 2.4 \times 10^{-13}$) was   obtained
\cite{Primero}.   Later,  a  detailed  asteroseismic  model  for
G117$-$B15A  was  computed \cite{NewA},  and  $m_{\rm  a} \cos  \beta <
4.4$~meV ($g_{\rm ae} < 1.2 \times 10^{-13}$) was obtained.
Subsequently, an upper limit of $26.5$~meV
($g_{\rm ae} < 7.4 \times 10^{-13}$) was inferred for the axion
mass using an improved asteroseismic
model for  G117$-$B15A, and  a better  treatment of  the uncertainties
involved    \cite{BKea08}.     Finally,     employing    the    latest
asteroseismic  model  \cite{2012MNRAS.420.1462R}  and  the  most
recent determination of  the rate of change of the  period at 215~s of
G117$-$B15A,  $\dot{\Pi}=   (4.19  \pm  0.73)  \times   10^{-15}$  s/s
\cite{2012ASPC..462..322K},      the      issue     was      revisited
\cite{2012MNRAS.424.2792C}.      The    215~s     period    in     the
asteroseismic model of G117$-$B15A was  associated with a $g$ mode
trapped in the  H envelope. The comparison of the  theoretical rate of
period  change associated  with the  asteroseismic model  with the
observed  one  suggested  the   existence  of  an  additional  cooling
mechanism in  this pulsating  white dwarf,  consistent with  axions of
mass $m_{\rm a} \cos \beta=  17.4$~meV ($g_{\rm ae}= 4.9 \times 10^{-13}$).

R548, that is, ZZ~Ceti itself
(the  prototype of  the  class), is  another DAV  star  that has  been
intensively studied for the last  few decades.  Since the discovery of
R548, there have  been multiple attempts to measure the  drift rate of
its pulsation  period at $\sim  213.13$~s, but only very  recently the
rate of change of this period with time has been measured for the very
first time.   Specifically, using  41 years of  time-series photometry
from  1970  to 2011  a  value  of  $\dot{\Pi}=  (3.3 \pm  1.1)  \times
10^{-15}$  s/s  was  obtained \cite{2013ApJ...771...17M}.   Using  the
measured  rate of  period change  of R548  and an  asteroseismic
model for this star \cite{2012MNRAS.420.1462R},  a value of $m_{\rm a}
\cos \beta=  17.1$ meV ($g_{\rm ae}= 4.8 \times 10^{-13}$)
was inferred  \cite{2012JCAP...12..010C}.  The
fact that  both inferences for  the axion mass  are so similar  is not
surprising, in  view of  that both G117$-$B15A  and R548  have similar
effective  temperatures, masses,  and  pulsation characteristics.   In
particular,  it turns  out that  the $k= 2,\ \ell=  1$ mode  lies at
nearly  identical  periods  for  these  stars  ---  $\sim  215$~s  for
G117$-$B15A and $\sim 213$~s for R548.

Nevertheless, a cautionary remark is  in order here. In particular, it
turns out  that the  upper limit  for the mass  of the  axion obtained
using the  white dwarf luminosity  function of the Galactic Disk
\cite{2014JCAP...10..069M} is in  conflict with  a value as
high as $m_{\rm  a} \cos  \beta \sim 17$~meV,  the  value  derived
from  emplopying  the  pulsating  stars
G117$-$B15A and  R548.  Indeed,  a detailed  fit of  the shape  of the
white dwarf luminosity function of the Galactic Disk at the
luminosities relevant for axion emission,  disfavor  axion
masses   $m_{\rm  a}  \cos  \beta  \gtrsim
10$~meV. However, it  is worth noting that the high  mass of the axion
derived using asteroseismic tools is a direct consequence of the
identification of the $\sim 215$~s  ($\sim 213$~s) mode of G117$-$B15A
(R548)     as    a     mode    trapped     in    the     H    envelope
\cite{2012MNRAS.424.2792C,2012JCAP...12..010C}.   The  rate of  period
change  of  a  trapped  mode  is  reduced  by  residual  gravitational
contraction, because  the oscillations  of trapped modes  are confined
close to the surface \cite{B96}.  Thus, it is of the utmost importance
to study  a pulsating white dwarf  star in which the  mode of interest
(one for which  the rate of period  change is measured) is  not a mode
trapped in the envelope, but  rather a mode with appreciable amplitude
throughout the star.

Recently, it has been possible to derive for the first time a value of
the rate  of change for two  periods in a third  ZZ~Ceti star, L19$-$2
\cite{2015ASPC..493..199S}.   This  star  (also  known  as  MY~Aps  or
WD1425$-$811) is  a relatively  bright pulsator, with  a spectroscopic
determination  of its  effective  temperature and  gravity of  $T_{\rm
  eff}= 12\, 100 \pm 200$~K and  $\log g= 8.21 \pm 0.10$, respectively
\cite{2004ApJ...600..404B,2000BaltA...9..119K}.    This   star   shows
oscillation periods $\Pi$ (and milli-modulation\footnote{One
  milli-modulation amplitude is the
  Fourier amplitude of a signal with a fractional intensity
  variation of 0.1 per cent.} amplitudes  $A$) of 113.8~s (2.05 mma),
118.7~s  (1.34  mma), 143.6~s  (0.49  mma),  192.6~s (5.92  mma),  and
350.2~s   (0.77   mma),   that  correspond   to   genuine   eigenmodes
\cite{1998BaltA...7..159S}.  The stability of the periods of the modes
at 192.6~s and 113.8~s (associated to the largest-amplitude modes) has
been recently  investigated \cite{2015ASPC..493..199S}. L19$-$2 is
a well studied star and has been
the     subject     of    several     asteroseismic     analyses
\cite{2001ApJ...552..326B,         2009MNRAS.396.1709C,         Rea12,
2013ApJ...779...58R}, for which we have reliable determinations of its
effective temperature  and surface gravity.  Moreover,  the results of
all  the  asteroseismic  models  agree  with  the  spectroscopic
inferences.  Thus, this  star provides an unique  opportunity to study
the role of axions in white dwarf cooling.

In  this paper,  we employ  a detailed  asteroseismic  model for
L19$-$2   which  is   an   improved  version   of   that  derived   in
Ref. \cite{2013ApJ...779...58R} and use the measurement of the rate of
change of  the 113.8~s and 192.6~s  periods to set  new constraints on
the mass of the axion.  The paper is organized as follows.  In Sect.~2
we give a  succinct account of the measurements of  the rate of period
change  in  L19$-$2,  while  in Sect.~\ref{asteroseismic}  we  briefly
present  our asteroseismic  model for  L19$-$2, and  describe in
detail some  propagation properties of  the pulsation modes.   In this
section  we also  briefly discuss  the uncertainties  involved  in our
analysis.  Sect.~\ref{axion_emission} follows, where we describe the
impact   that  the   inclusion  of   the  axion   emissivity   in  the
asteroseismic    model    has    on   the    pulsation    modes.
Sect.~\ref{axion_mass} is  devoted to place an  improved constraint on
the axion mass.  Finally,  in Sect.~\ref{conclusions} we summarize our
findings and we present our concluding remarks.


\section{Measurements of $\dot{\Pi}$ for the periods at 113 s and 192 s}
\label{observations}

\begin{table}
\centering
\caption{Characteristics  of   L19$-$2  obtained   using  spectroscopy
  \cite{2000BaltA...9..119K,2004ApJ...600..404B} and  according to the
  asteroseismic  model  \cite{2013ApJ...779...58R}.  The  quoted
  uncertainties  in the  asteroseismic  model  are the  internal
  errors of the period-fit procedure.}
\begin{tabular}{lcc}
\hline
\hline
 Quantity                        & Spectroscopy      & Asteroseismology            \\
\hline
\hline
$T_{\rm eff}$ [K]                & $12\,100 \pm 200$ & $11\,980 \pm 300$          \\
$M_{\star}/M_{\sun}$                  &  $0.727\pm 0.062$    & $0.705 \pm 0.023$  \\
$\log g$                         & $8.21\pm 0.10$    & $8.177 \pm 0.088$        \\
$\log(R_*/R_{\odot})$            &    ---            & $-1.945\pm 0.037$         \\   
$\log(L_*/L_{\odot})$            &    ---            & $-2.622\pm0.046$     \\
$M_{\rm He}/M_{\star}$                 &    ---            & $7.63 \times 10^{-3}$   \\
$M_{\rm H}/M_{\star}$                  &    ---            & $3.14 \times 10^{-5}$   \\
$X_{\rm C},X_{\rm O}$ (center)   &    ---             & $0.326, 0.661$ \\
\hline
\hline
\end{tabular}\\
\label{table1}
\end{table}

L19$-$2 was found  to be a pulsating white dwarf  of the ZZ~Ceti class
in  1977 \cite{1977ApJ...214L.123M}.   Subsequently, it  was monitored
during  nearly   five  years  from  the   South  African  Astronomical
Observatory (SAAO) \cite{1982MNRAS.200..563O,1987MNRAS.228..949O}.  It
was found that  the star pulsates with at least  five periods at $\sim
350$, $\sim 192$,  $\sim 143$, $\sim 118$ and $\sim  113$~s.  Also, it
was possible  to study the stability  of the modes with  periods $\sim
113$ and $\sim  192$~s, and upper limits to the  rate of period change
of  $\sim 2  \times 10^{-14}$~s/s  were placed  for both  periods. The
measurement of $\dot{\Pi}$ in a pulsating white dwarf is a challenging
observational endeavour,  since a  large photometric  data set  with a
very  long   observational  time   baseline  (typically   decades)  is
necessary.  Fortunately,  L19$-$2 was  the  target  of a  Whole  Earth
Telescope   (WET)   \cite{1990ApJ...361..309N}    campaign   in   1995
\cite{1998BaltA...7..159S} employing only three southern
observatories. In addition, single-site observations covering the
interval 1994 to 2013 have been obtained using the one metre
telescope at the University of Canterbury Mt John Observatory (UCMJO).
A comprehensive analysis of the 1995 WET data along with single
site  observations  will be  presented  in  a forthcoming  publication
\cite{SC16},   but    a   summary    has   already    been   presented
\cite{2015ASPC..493..199S}.  The stability  of  the  periods at  $\sim
192$~s and $\sim 113$~s was investigated,  and a rate of period change
of $\dot{\Pi}= (4.0  \pm 0.6) \times 10^{-15}$~s/s for  both modes was
found.  By taking into account a  contribution due to proper motion of
the star \cite{1995A&A...295L..17P} of  $\sim 1.0 \times 10^{-15}$~s/s
\cite{2015ASPC..493..199S}, we obtain $\dot{\Pi}= (3.0 \pm 0.6) \times
10^{-15}$~s/s. This  is the  value of $\dot{\Pi}$  for the  periods at
$\sim 113$~s and $\sim 192$~s that we will adopt in our analysis.


\section{Asteroseismic model for L19$-$2}
\label{asteroseismic}

Here, we  describe briefly the asteroseismic  model for L19$-$2,
and   refer  the   interested   reader  to   a  previous   publication
\cite{2013ApJ...779...58R}   for   further  details   concerning   the
asteroseismic approach  adopted. A  detailed asteroseismic
analysis of L19$-$2 using a grid of DA white dwarf evolutionary models
characterized by chemical profiles  for both the  core and
the envelope consistent with the evolutionary history of the
progenitor stars, and covering a wide range of stellar masses,
thicknesses
of the  hydrogen envelope and  effective temperatures was  carried out
\cite{2013ApJ...779...58R}.  These models were computed employing the
{\tt  LPCODE}  evolutionary  code  \cite{Altea05}.   The  evolutionary
calculations   were   carried  out   from   the   ZAMS,  through   the
thermally-pulsing and mass-loss phases on  the AGB, and finally to the
domain  of   planetary  nebulae  and  white   dwarfs.   The  effective
temperature, the stellar mass and the mass of the H envelope of our DA
white dwarf models  are within the intervals $14\,000  \ga T_{\rm eff}
\ga 9\,000$~K, $0.525 \la M_{\star} \la 1.050 M_{\sun}$, and $-9.4 \la
\log(M_{\rm H}/M_{\star})  \la -3.6$, respectively.  The  value of the
upper limit of $M_{\rm H}$ is dependent on $M_{\star}$ and is given by
the previous  evolution of the  progenitor star.  For  simplicity, the
mass of He  was kept fixed at the value  predicted by the evolutionary
computations  for   each  sequence.    Different  treatments   of  the
crystallization process  were taken into account  self-consistently in
the  evolution   and  pulsations   of  our   DA  white   dwarf  models
\cite{2013ApJ...779...58R}.

\begin{table}
\centering
\caption{The  observed   and  theoretical   asteroseismic  model
  periods  and  rates of  period  change.   The  capital letter  T  in
  parenthesis for  the mode  with $k= 1,\ \ell=  1$  stands for  a mode
  trapped in the H envelope.}
\begin{tabular}{cccccc}
\hline
\hline
$\Pi^{\rm o}$        &  $\Pi^{\rm t}$      &  $\dot{\Pi}^{\rm o}$  & $\dot{\Pi}^{\rm t}$  & $k$ & $\ell$ \\
\noalign{\smallskip}                     
$[$s$]$              &   $[$s$]$        &  $[10^{-15} $s/s$]$   &  $[10^{-15}  $s/s$]$ &        &     \\
\hline
 118.7              &    117.21 (T)     &  ---                  &  0.47           &   1    &  1  \\ 
 192.6              &    192.88         &  $3.0 \pm  0.6$       &  2.41              &   2    &  1  \\  
 113.8              &    113.41         &  $3.0 \pm  0.6$       &  1.42              &   2    &  2  \\
 143.6              &    145.41         &  ---                  &  0.89              &   3    &  2  \\ 
\hline
\hline
\end{tabular}
\label{table2}
\end{table}

In order to  find an asteroseismic model for  L19$-$2, the model
that minimizes  a quality function  that measures the goodness  of the
fit of the theoretical ($\Pi^{\rm t}$) to the observed ($\Pi^{\rm o}$)
periods  was   sought  \cite{2013ApJ...779...58R}.    The  theoretical
periods were  assessed by means of  the adiabatic version of  the {\tt
  LP-PUL} pulsation  code \cite{CA06}.   A single best-fit  model with
the  characteristics   shown  in  Table~\ref{table1}  was   found.  
This model is an improved version of the  asteroseismic model computed 
previously for L19$-$2 \cite{2013ApJ...779...58R}. We
identified the modes $118.7$~s $(k= 1,\ \ell= 1)$,
$192.6$~s $(k= 2,\ \ell= 1)$, $\Pi^{\rm o}= 113.8$~s $(k= 2,\ \ell= 2)$,
and $143.4$~s $(k= 3,\ \ell= 2)$. Within the scope of our set of
pulsation models, 
these were the only possible identifications. The  second column  of
Table~\ref{table1}   contains  the   $T_{\rm  eff}$,   $\log  g$   and
$M_{\star}$   of   L19$-$2    according   to   spectroscopic   studies
\cite{2004ApJ...600..404B,   2000BaltA...9..119K}.    The   parameters
characterizing the asteroseismic model are shown in column 3.

In Table~\ref{table2} we compare  the observed and theoretical periods
and  rates of  change  of the  periods.   As can  be  seen, the  model
reproduces very well  the observed periods. This  is particularly true
for  the case  of  the 113.8~s  and 192.6~s  periods.   Note that  the
observed rate  of change of  the 113.8~s  and 192.6~s periods  is more
than 2  and 1.25  times larger,  respectively, than  the theoretically
expected values.  If we assume that the rate of period change of these
two modes  reflects the evolutionary  timescale of the star,  then the
disagreement between the observed  and theoretical values of $\dot\Pi$
would be a hint that L19$-$2 could be cooling faster than predicted by
the standard theory of white dwarf evolution.

\begin{figure} 
\begin{center}
\includegraphics[clip,width=0.9\columnwidth]{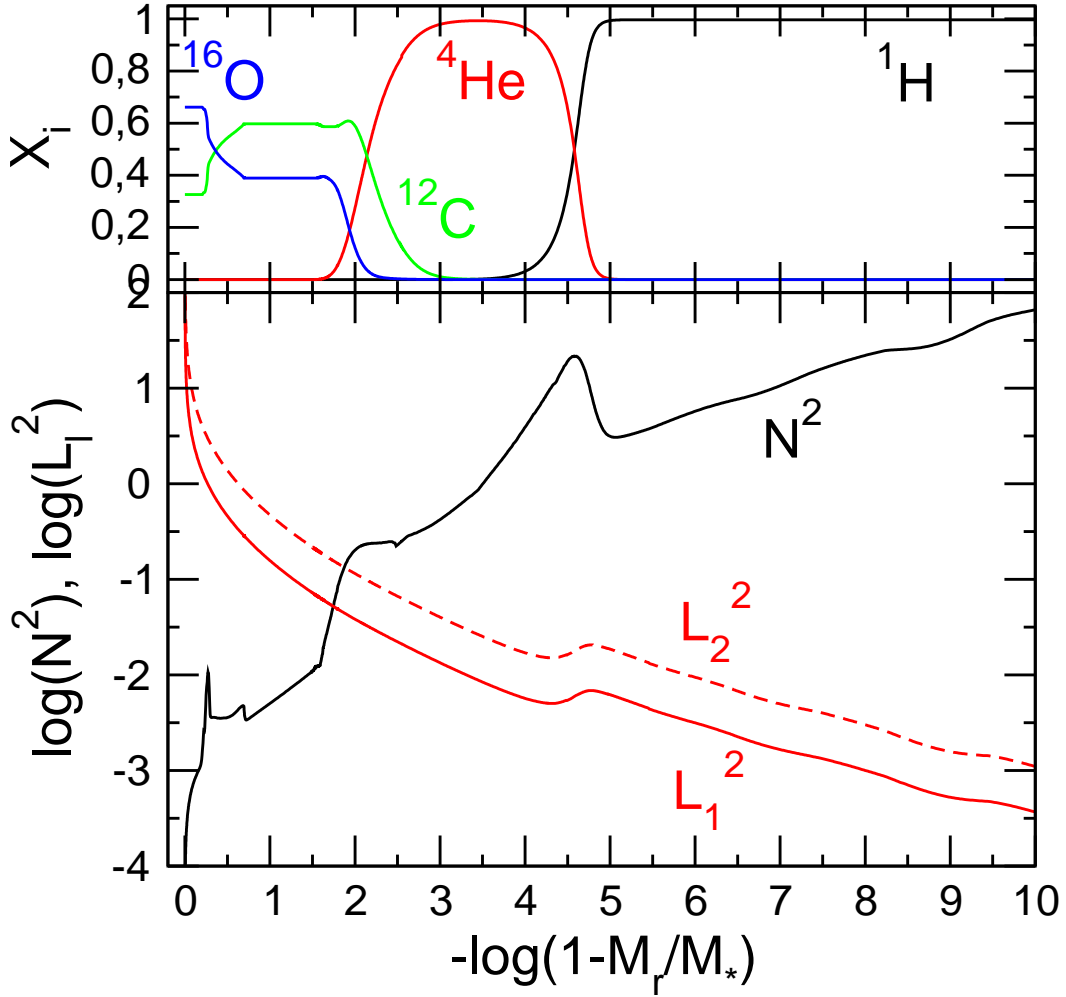} 
\caption{The  internal chemical  stratification  (upper  panel) and  a
  propagation diagram  (lower panel) of our  asteroseismic model
  for  L19$-$2,  characterized  by  $M_{\star}=  0.705  M_{\sun}$  and
  $T_{\rm eff}= 12\,033$~K.}
\label{figure1} 
\end{center}
\end{figure}

The internal  chemical stratification and the  propagation diagram ---
the run  of the  critical frequencies, namely  the Brunt-V\"ais\"al\"a
frequency, $N$,  and the Lamb  frequency, $L_{\ell}$ for $\ell=  1, 2$
--- of our asteroseismic model  are shown in Fig.~\ref{figure1}.
Each  chemical  transition  region   produces  clear  and  distinctive
features  in  $N$,  which  are eventually  responsible  for  the  mode
trapping  properties of  the model  \cite{B96,2002A&A...387..531C}. In
the core of the star, there is a peak at $-\log(q) \approx 0.3$ (at $q
\equiv 1-M_r/M_{\star}$) resulting from  steep variations in the inner
CO profile which are caused by the occurrence of extra mixing episodes
beyond the fully  convective core during central  helium burning.  The
apparent bump  in $N^2$  at $-\log(q)  \approx 2-3$  is caused  by the
chemical transition of He to C and O resulting from nuclear processing
in   the  previous   AGB   and   thermally-pulsing  AGB   evolutionary
phases. Finally,  there is the transition  region between H and  He at
$-\log(q)  \approx 4.5$,  which is  smoothly shaped  by the  action of
time-dependent element diffusion.

\subsection{Mode trapping}
\label{propa}

\begin{figure} 
\begin{center}
\includegraphics[clip,width=1\columnwidth]{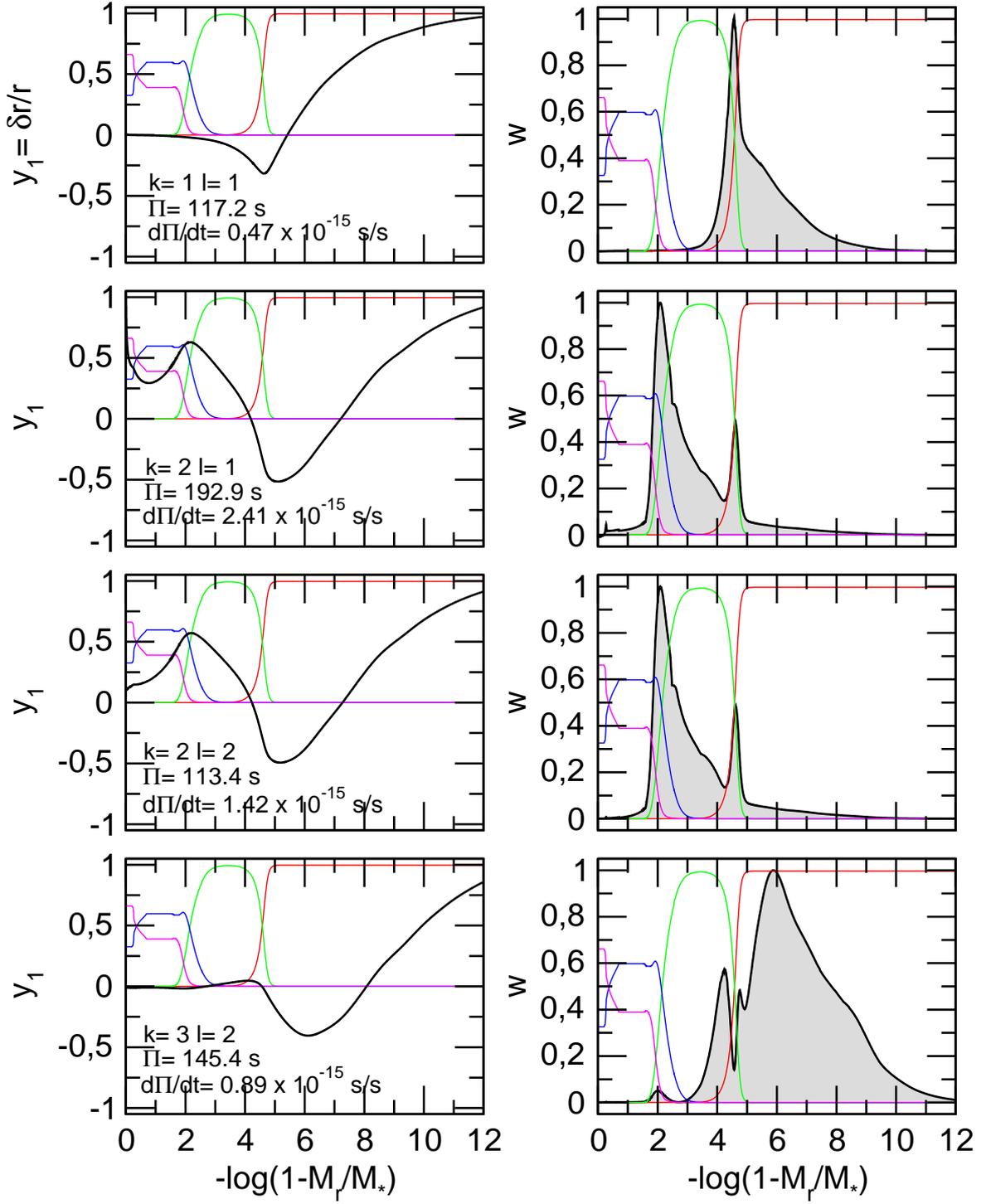} 
\caption{The radial eigenfunction $y_1=\xi_r/r$  (left panels) and the
  normalized weight  function $w$ (right  panels, shaded) in  terms of
  the outer mass fraction, corresponding to the modes with $(k, \ell)=
  (1,1),  (2,1),  (2,2)$, and  $(3,2)$  (from  top  to bottom)  of  the
  asteroseismic model.  The chemical  abundances are also shown,
  to provide a reference.}
\label{figure2} 
\end{center}
\end{figure}

Table~\ref{table2} also  reveals that the  rates of period  change for
the various  modes are substantially  different.  This is  because the
modes have distinct mode  trapping properties.  Briefly, mode trapping
is a mechanical resonance between  the local oscillation wavelength of
a pulsation mode with the thickness of one of the compositional layers
(for instance,  the H envelope or  the He buffer).  Mode  trapping can
reduce the rate of  period change of a mode by up to  a factor of 2 if
it  is trapped  in  the  outer H  envelope  \cite{B96}.  The  trapping
properties of pulsation modes can  be studied \cite{CA06} by examining
their radial eigenfunction ($y_1=  \xi_r/r$) and weight function, $w$.
Fig.~\ref{figure2} shows these functions for the modes with $(k, \ell)=
(1,1)$, (2,1), (2,2), and $(3,2)$  of the asteroseismic model of
L19$-$2.  The  weight function  of a  given mode  allows one to  infer the
regions  of the  star that  most  contribute to  the period  formation
\cite{1985ApJ...295..547K}.  Clearly, $y_1$ and  $w$ for the mode with
$k= 1,\ \ell= 1$ have appreciable amplitudes only in the region bounded
by the  He/H interface and  the stellar surface, and  so it is  a mode
strongly trapped in the outer H envelope. This property for the $k= 1,\
\ell= 1$ mode holds also for all the models with structural parameters
($M_{\star}, M_{\rm H}, T_{\rm eff}$) similar to those of the best-fit
model.  Since the region where this  mode is relevant is located close
to the  surface, gravitational contraction (that  is still appreciable
in these regions) acts reducing the period change due to cooling alone
\cite{B96}. This  is the reason why  the rate of period  change of the
$k= 1,\ \ell= 1$ ($\Pi \sim  118$~s) mode is small.   In contrast, the
non-trapped modes with  $k= 2,\ \ell= 2$ ($\Pi \sim  113$~s) and $k= 2,\
\ell= 1$  ($\Pi \sim 192$~s) have amplitudes below  the He/H interface
and, in particular, at the He/C/O chemical transition region. Finally,
the mode  with $k= 3,\ \ell= 2$  ($\Pi \sim 143$~s)  has mode-trapping
properties which  are intermediate  between those  of the  trapped and
non-trapped  modes.  Since  non-trapped  modes  are  not  affected  by
gravitational contraction, they have larger rates of period change.

In summary, the rates  of period change for the
$k= 2,\ \ell= 2$ ($\Pi \sim  113$~s) and
$k= 2,\ \ell= 1$ ($\Pi \sim  192$~s) modes  are not
affected  by   gravitational  contraction,  implying  that   they  are
sensitive to cooling and are useful for our analysis.
In addition, given the observed characteristics of the L19$-$2
pulsation spectrum, these two modes are the only suitable ones for
investigating long term period changes.

We point that, despite the fact that the asteroseismic  best-fit
  model reproduces the observed periods quite well (see Table \ref{table2}),
  the predicted
  $\dot{\Pi}$ values for the 192.6 s and 113.8 s modes differ by a
  factor $(2.41/1.42) \sim 1.7$. On the other hand, taking into account the
  observational $\dot{\Pi}$ values and error bars, this factor is
  actually $\frac{(3.0+0.6)}{(3.0-0.6)}= 1.5$. In other words, our
  asteroseismic model does not match the observed rate of period
  change of both modes simultaneously. In the next
  section, we discuss some probable sources of uncertainty affecting
  the theoretical rates of period change.


\subsection{Uncertainties in the theoretical value of $\dot{\Pi}$}
\label{errors}

Several sources of uncertainty affect the value of the rates of period
change  for the  $k= 2,\ \ell= 1$  and $k= 2,\ \ell= 2$ modes  of our
asteroseismic model.  This is  a crucial  point to  estimate the
uncertainties in the derived axion mass.  We consider two main sources
of    errors.     The    first    one    is     the    poorly    known
$^{12}$C$(\alpha,\gamma)^{16}$O  nuclear  reaction rate at relevant
stellar energies,  whereas  the
second  one stems  from the  uncertainties  in the  parameters of  the
asteroseismic  model ($M_{\star}$,  $T_{\rm  eff}$, and  $M_{\rm H}$).

The $^{12}$C$(\alpha,\gamma)^{16}$O  nuclear reaction plays  a crucial
role in shaping  the chemical structure of the cores  of white dwarfs.
Indeed, the final  carbon-oxygen stratification of a  newly born white
dwarf strongly depends on the  efficiency of this reaction rate during
the late stages of core helium burning \cite{2010ApJ...717..897A}.  In
our evolutionary calculations, we employed the most commonly used rate
for          the         $^{12}$C$(\alpha,\gamma)^{16}$O          rate
\cite{1999NuPhA.656....3A}.  Unfortunately, the  rate of this reaction
is not accurately known. Estimated uncertainties of this reaction rate
have been compiled \cite{2002ApJ...567..643K}.  The uncertainties vary
from a factor $\sim 1.4$ to  a factor $\sim 0.6$. The uncertainties of
the  theoretical rates  of  period change  of the  asteroseismic
model of the ZZ Ceti star  G117$-$B15A due to the uncertainties in the
$^{12}$C$(\alpha,\gamma)^{16}$O reaction rate have also been estimated
\cite{2012MNRAS.424.2792C}.    Not   surprisingly,    the   value   of
$\dot{\Pi}$ for  a mode trapped  in the  H envelope exhibits  a modest
variation,  of about  $2.5 \%$,  when the  central $^{16}$O  abundance
strongly  increases from  $X_{^{16}{\rm O}}=  0.482$ to  $X_{^{16}{\rm
O}}= 0.795$.  At variance with this, in the case of non-trapped modes,
the rate  of period change experiences  quite large changes, of  up to
$\sim 60 \%$.  These results allow  us to assess in an approximate way
the impact of the uncertainties of the $^{12}$C$(\alpha,\gamma)^{16}$O
reaction  rate on  the  value of  the  rate of  period  change of  the
non-trapped modes of interest in L19$-$2, that is, the $k= 2,\ \ell= 1$
and $k= 2,\ \ell=  2$ modes.  We  adopt an  uncertainty $\varepsilon_1
\sim  0.85  \times  10^{-15}$~s/s  ($\varepsilon_1  \sim  1.45  \times
10^{-15}$~s/s) for  the $\dot{\Pi}^{\rm  t}$ associated with  the period
$\sim 113$~s ($\sim 192$~s).

Another source of error in the theoretical values of $\dot{\Pi}$ comes
from the  uncertainties in the asteroseismic  model for L19$-$2.
 These are \emph{internal} errors of the period-fit procedure
\cite{2012MNRAS.424.2792C}. We estimate that the uncertainty in  the
rate of period change for the
modes with $k= 2,\ \ell= 1$ and  $k= 2,\ \ell= 2$ is $\varepsilon_2 \sim
0.06 \times 10^{-15}$~s/s at most \cite{2012MNRAS.424.2792C}.  This is
a small uncertainty that is completely negligible when compared to the
uncertainties      arising       from      the       poorly      known
$^{12}$C$(\alpha,\gamma)^{16}$O reaction rate.

In addition to the $^{12}$C$(\alpha,\gamma)^{16}$O reaction rate and
the period-fit procedure, other possible sources of uncertainty
are the amount of mixing that occurs during both the core He burning and
thermally pulsing AGB phases, the metallicity of the white dwarf
progenitor, rotation, etc. We caution that these sources of uncertainty
(which have been neglected in our treatment) could affect to some extent
the values of the theoretical rates of period change. The precise
assessment of the impact of these uncertainties on $\dot{\Pi}$
is beyond the scope of the present paper but will be addressed in
a future publication.


\section{Impact of axion emission on the rates of period change}
\label{axion_emission}

In the previous section we have presented results of periods and rates
of  period change  for L19$-$2  that do  not take  into account
energy sources other than gravothermal energy for the evolutionary
cooling of the  star.  We  have found  that the  theoretically
expected  rates of
period change of  the modes with $k= 2,\ \ell=  1$ and $k= 2,\ \ell= 2$
are smaller  than the  $\dot{\Pi}$ value measured  for the  periods at
$192$~s  and $113$~s  of  L19$-$2, suggesting  the  existence of  some
additional cooling mechanism in this star.  Hereafter, we shall assume
that  this  additional  cooling  can be  entirely  attributed  to  the
emission of axions.

\begin{figure} 
\begin{center}
\includegraphics[clip,width=0.9\columnwidth]{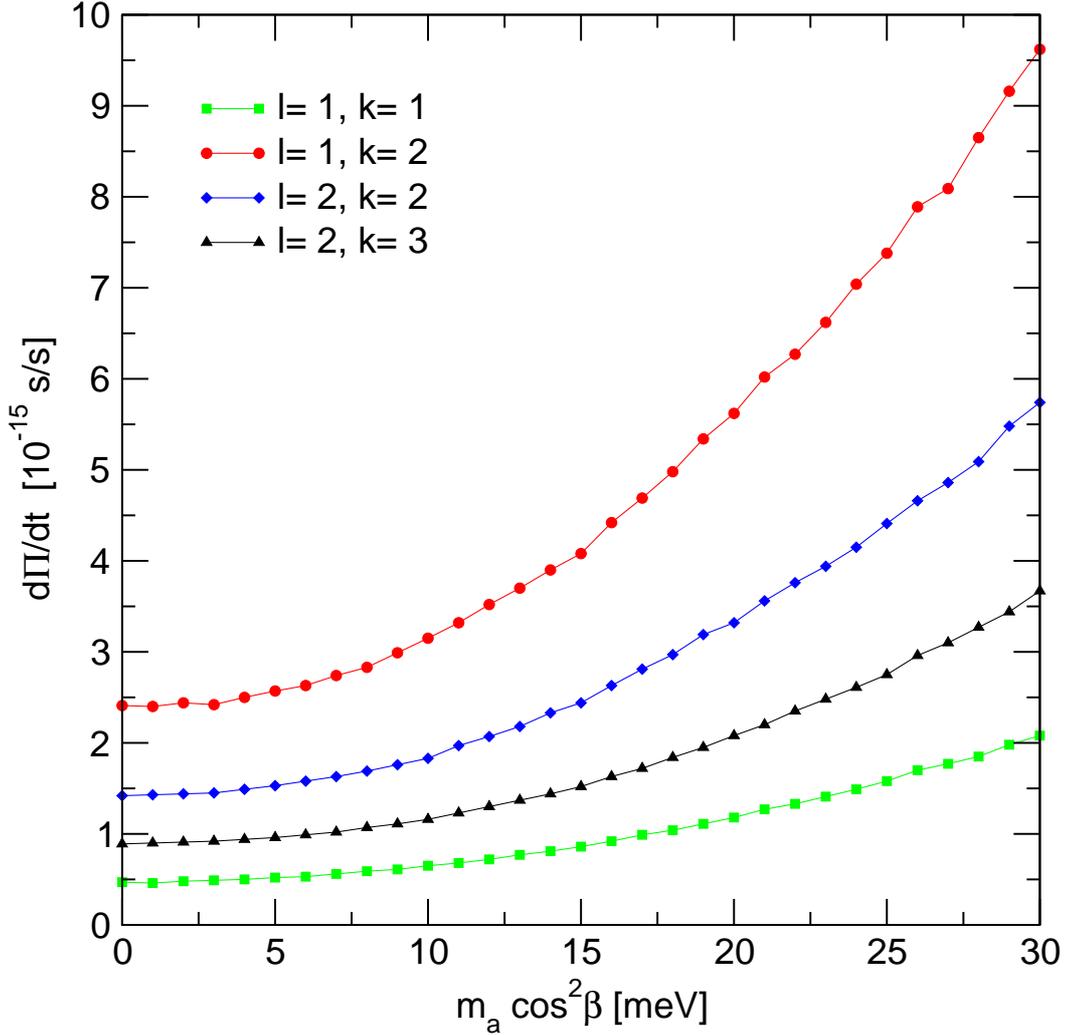} 
\caption{The rates of period change of  the modes with
  $k=1, 2,\ \ell= 1$
  and $k= 2, 3,\ \ell= 2$, corresponding to our asteroseismic
  model for L19$-$2 in terms of the axion mass.}
\label{figure3} 
\end{center}
\end{figure}

We computed  a set of  DA white dwarf cooling  sequences incorporating
axion         emission        \cite{NewA,         2012MNRAS.424.2792C,
2012JCAP...12..010C}.   This  was  done  considering  different  axion
masses and  the same structural parameters  ($M_{\star}$, $M_{\rm H}$)
as those  of the  asteroseismic model discussed above. We
adopted a  range of
values for  the mass of the  axion $0\leq m_{\rm a}  \cos^2 \beta \leq
30$~meV with  a step size $\Delta m_{\rm  a}= 1$~meV.
In all  the cases we
employed  the most  reliable axion  emission rates  \cite{Nea87}.  The
evolutionary  calculations  including  the  emission  of  axions  were
started at evolutionary stages long before the ZZ~Ceti phase to ensure
that the cumulative  effects of axion emission  reached an equilibrium
value before the theoretical models  have the effective temperature of
L19$-$2.

We computed  the pulsation periods and  the rates of period  change of
the  asteroseismic  model  ($T_{\rm eff}  \sim  12\,030$~K)  for
increasing  values of  $m_{\rm a}$.  The variation  in the  periods is
negligible,  in  spite  of  the  rather wide  range  of  axion  masses
considered.  This is  an  important property  as previously noted
\cite{NewA}, that allows one to put constraints  on the mass of the axion.
The rates of period change for the  modes with $k= 1, 2,\ \ell= 1$, and
$k= 2, 3,\ \ell= 2$ of the  asteroseismic model  for increasing
values of  $m_{\rm a}$ are  displayed in Fig.~\ref{figure3}.   At odds
with  what  happens   with  the  pulsation  periods,   the  values  of
$\dot{\Pi}^{\rm t}$  are strongly  affected by the  additional cooling
source, substantially increasing for increasing values of $m_{\rm a}$.
In particular,  the rate  of period  change of  the modes  with
$k=2,\ \ell= 1$ and  $k= 2,\ \ell= 2$, which  are the  relevant ones  in the
present analysis,  increase by a  factor of about  4 for the  range of
axion masses considered here.

\section{The axion mass}
\label{axion_mass}

\begin{figure} 
\begin{center}
\includegraphics[clip,width=0.9\columnwidth]{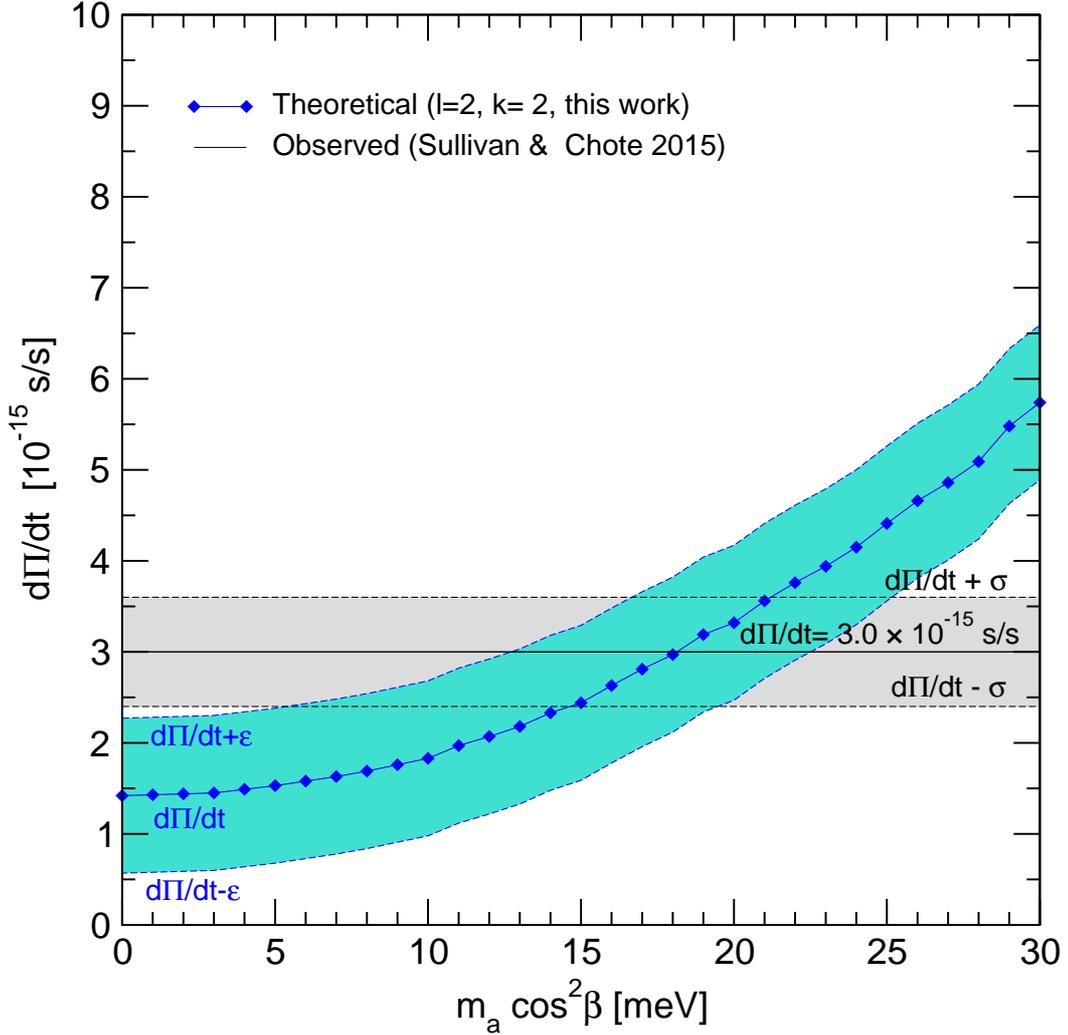} 
\caption{The rate of period change for the mode with $k= 2,\ \ell= 2$
  ($\Pi \sim 113$ s) of our  asteroseismic model in terms of the
  axion mass  (solid blue  curve with  diamonds).  Dashed  blue curves
  represent  the   errors  in   $\dot{\Pi}$  due  to   {\sl  internal}
  uncertainties  in  the  modeling   and  in  the  asteroseismic
  procedure  ($\varepsilon  \sim   0.85  \times  10^{-15}$~s/s).   The
  horizontal solid  black line  indicates the  observed value  and the
  dashed     black    lines     its    corresponding     uncertainties
  \cite{2015ASPC..493..199S}.}
\label{figure4} 
\end{center}
\end{figure}

\begin{figure} 
\begin{center}
\includegraphics[clip,width=0.9\columnwidth]{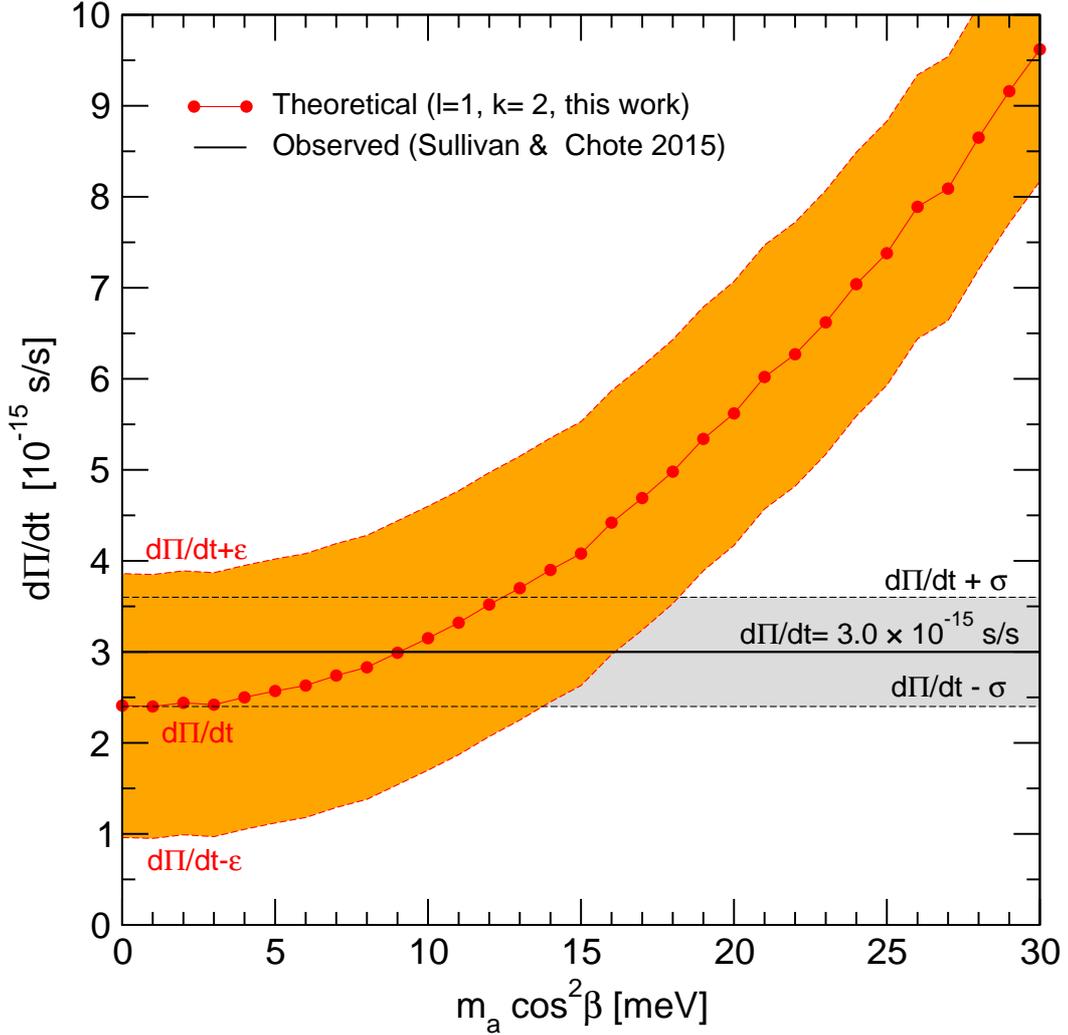} 
\caption{Similar to Fig.  \ref{figure4}, but for the  mode with
  $k= 2,\ \ell= 1$  ($\Pi  \sim  192$ s).  In  this  case, the  theoretical
  uncertainty is  of the  rate of period  change is  $\varepsilon \sim
  1.45 \times 10^{-15}$~s/s.}
  \label{figure5} 
\end{center}
\end{figure}

Here, we focus on the modes with $k=2,\ \ell=  1$ and
$k= 2,\ \ell= 2$,
for which  we have a measurement  of their rate of  period change.  In
Fig.~\ref{figure4}  we  show  the  theoretical  value  of  $\dot{\Pi}$
corresponding to the period $\Pi  \sim 113$~s for increasing values of
the axion  mass (blue solid  curve).  The dashed curves  embracing the
solid  curve represent  the estimated uncertainty  in the
theoretical value  of
$\dot{\Pi}$, $\varepsilon_{\dot{\Pi}}= 0.85 \times 10^{-15}$~s/s. This
value was obtained considering the  uncertainty introduced by our lack
of precise  knowledge of the  $^{12}$C$(\alpha,\gamma)^{16}$O reaction
rate ($\varepsilon_1 \sim 0.85 \times 10^{-15}$~s/s), which completely
dominates   the   errors   in   the  asteroseismic   model
($\varepsilon_2     \sim     0.06    \times     10^{-15}$~s/s,     see
Sect. \ref{errors}).   We are  assuming that  the uncertainty  for the
case in which $m_{\rm  a} > 0$ is the same as  that computed for the
case in which  $m_{\rm a}= 0$.  If we consider  one standard deviation
from the  observational value, we obtain  an upper limit on  the axion
mass of  $m_{\rm a} \cos^2  \beta \lesssim 25$~meV
($g_{\rm ae}= 7 \times 10^{-13}$). If,  instead, we
consider the mode with period $\Pi  \sim 192$~s, we obtain the results
displayed  in  Fig.~\ref{figure5}. In  this  case,  the dashed  curves
embracing  the  solid  red  curve represent  the  uncertainty  in  the
theoretical  value  of   $\dot{\Pi}$,  $\varepsilon_{\dot{\Pi}}=  1.45
\times 10^{-15}$~s/s  (Sect. \ref{errors}).  We obtain  an upper limit
for the axion mass of $m_{\rm a} \cos^2 \beta \lesssim 18$ meV
($g_{\rm ae}= 5.1 \times 10^{-13}$).

In  summary, based  on  the rate  of period  change  measured for  the
ZZ~Ceti star  L19$-$2, we derive an  upper limit of $m_{\rm  a} \cos^2
\beta$ of $25.1$~meV ($g_{\rm ae}= 7.0 \times 10^{-13}$).
We note that, if the  uncertainties of the
theoretical rates of period change  were drastically reduced, then the
upper limit on the axion mass would be lowered ($\sim 13-21$~meV).

\section{Discussion and conclusions}
\label{conclusions}

In this  paper we have derived  new bounds for  the mass of the
hypothetical  particle called axion.  For this purpose we  used a
detailed asteroseismic  model for L19$-$2, a  hot and relatively
massive ZZ Ceti star, derived  from fully evolutionary calculations of
DA white dwarfs \cite{2013ApJ...779...58R}, and we employed the recent
determination of the  rate of period change  for the largest-amplitude
modes ($k= 2,\ \ell= 2$, $\Pi \approx 113$~s and  $k=2,\ \ell= 1$,
$\Pi \approx 192$~s) of this star \cite{2015ASPC..493..199S}.

We first compared the observed rate  of period change for these modes,
$\dot{\Pi}=  (3.0 \pm  0.6) \times  10^{-15}$~s/s, with  the rates  of
period change  of our  asteroseismic model,  $\dot{\Pi}^{\rm t}=
(1.42  \pm  0.85)  \times  10^{-15}$~s/s ($k= 2,\ \ell=  2$)  and
$\dot{\Pi}^{\rm t}=  (2.41 \pm 1.45) \times  10^{-15}$~s/s
($k= 2,\ \ell= 1$), computed  under the assumption that the cooling
of this star is governed only by the release  of
gravothermal energy. The fact that
the  observed  value  is  about $1.25-2$  times  larger  than  the
theoretically expected ones hints at the possibility
that  this star might be cooling faster
than the standard  theory of white dwarf evolution  predicts.
We took
into account the possible sources  of uncertainties in the theoretical
values of the rate of  period change.  The uncertainties affecting the
poorly determined $^{12}$C$(\alpha,\gamma)^{16}$O reaction rate have a
strong impact on  the theoretical value of the rates  of period change
for the two modes of interest, because these modes have non-negligible
amplitudes at the core regions of the asteroseismic model. These
uncertainties  largely  overcome  the internal  uncertainties  of  the
asteroseismic  model.   We  reached  the  conclusion   that  the
uncertainties in the  theoretical rates of period change  are of $\sim
60 \%$.

Next, we assumed that the  additional cooling necessary to account for
the  large  observed  rate  of  period  change  of  L19$-$2  could  be
attributed to axion  emission.  Then, we introduced  axion emission in
our asteroseismic model, considering  a sufficiently large range
of axion masses (between $0$ and  $30$~meV). We found that the periods
do not change, but the rates of period change are strongly affected by
the precise value of the axion mass.   We found that the mass of axion
necessary to account  for the observed rate of period  change could be
as high  as $\sim 18$~meV or  $\sim 25$~meV, depending on  the mode we
use in our analysis.  Specifically, when the mode with period
$\sim 113$~s was  employed, we obtained
$5 \lesssim m_{\rm a}  \cos^2 \beta \lesssim  25$, or, in terms
of the axion-electron coupling constant,
$1.4 < g_{\rm ae}/10^{-13}  < 7$
(Fig.  \ref{figure4}).  In  the  case of  the mode  with
period at $\sim 192$~s we derived an upper bound for $m_{\rm a} \cos^2
\beta$ of $\sim 18$~meV ($g_{\rm ae} \sim 5 \times 10^{-13}$),
although a  zero mass for the axion cannot
be discarded (see Fig. \ref{figure5}).

\begin{table}
\centering
\caption{Summary of the  different constraints on the  DFSZ axion mass
  from white dwarf stars.}
\begin{tabular}{cccc}
\hline
\hline
Method               & Object (class)      &     $m_{\rm a} \cos^2 \beta$ [meV]  & Reference \\
\noalign{\smallskip}
\hline
Asteroseismology     & G117$-$B15A (DAV)     & $\lesssim 10$    & \cite{Primero} \\     
Asteroseismology     & G117$-$B15A (DAV)     & $\lesssim 4.4$   & \cite{NewA}\\
Asteroseismology     & G117$-$B15A (DAV)     & $\lesssim 26.5$  & \cite{BKea08} \\
Asteroseismology     & G117$-$B15A (T, DAV)  & $14.7-19.7$      & \cite{2012MNRAS.424.2792C}\\
Asteroseismology     & R548 (T, DAV)         & $11.3-21.4$      & \cite{2012JCAP...12..010C}\\
Asteroseismology     & PG 1351+489 (NT, DBV) & $\lesssim 19.5$  & \cite{Battich}\\
Asteroseismology     & L19$-$2 (NT, DAV)     & $\lesssim 25$ & This work\\
\hline
WDLF                 & Observed WDLF         & $\lesssim 5$     & \cite{WDLF}\\
WDLF                 & Observed WDLF         & $\lesssim 6$     & \cite{2009JPhCS.172a2005I} \\ 
WDLF                 & Observed WDLF         & $\lesssim 10$    & \cite{2014JCAP...10..069M} \\ 
\hline
\hline
\end{tabular}
\label{table3}
\end{table}

\begin{figure} 
\begin{center}
\includegraphics[clip,width=1.0\columnwidth]{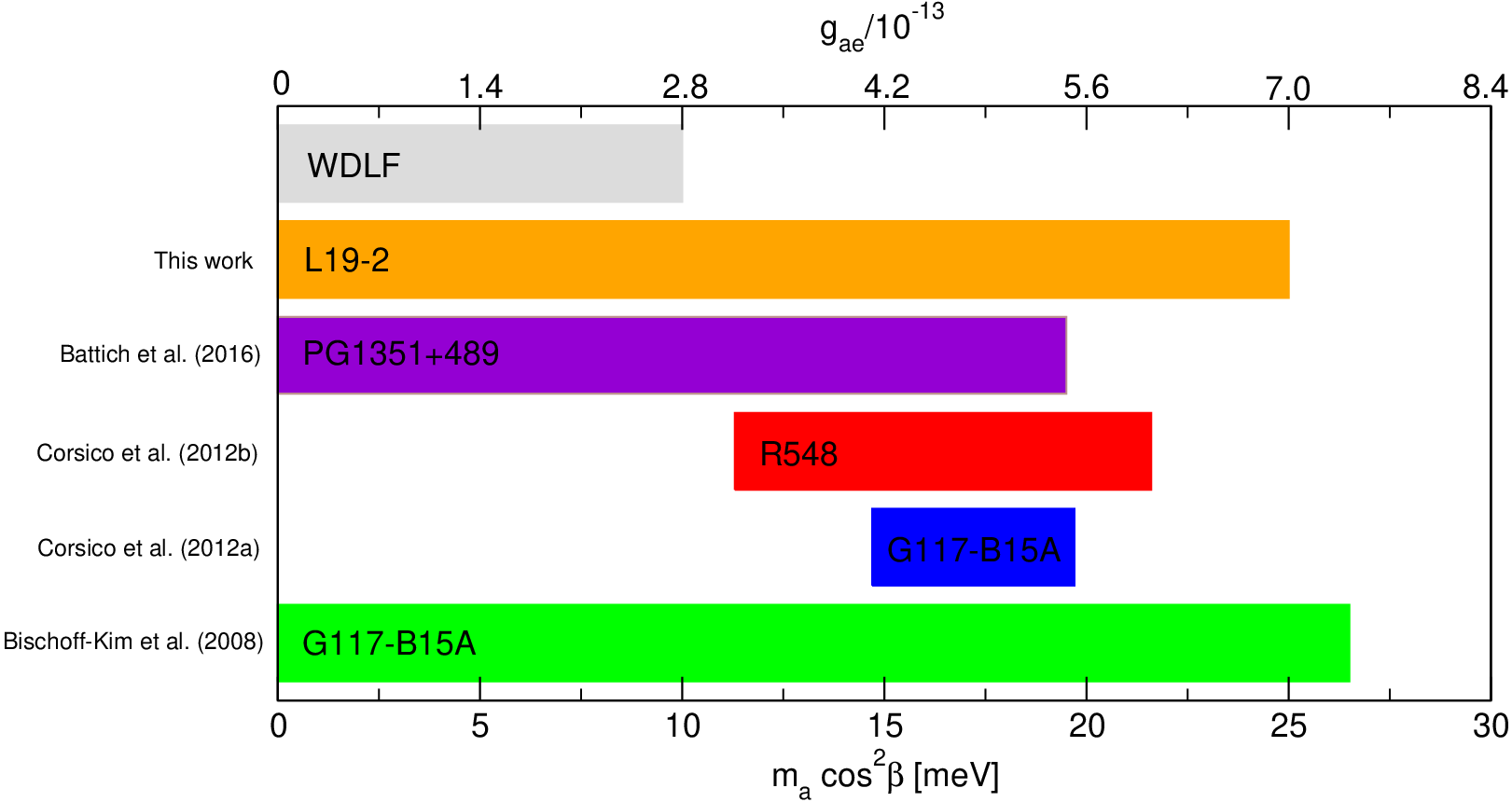} 
\caption{The recent asteroseismic inferences of the axion mass
  \cite{BKea08,2012MNRAS.424.2792C,2012JCAP...12..010C,Battich} 
  and the
  derivations of $m_{\rm a}$ employing the white  dwarf luminosity
  function \cite{WDLF,2009JPhCS.172a2005I,2014JCAP...10..069M}  (see Table \ref{table3}).} 
  \label{figure6} 
\end{center}
\end{figure} 

In  Table \ref{table3}  we summarize  the various  limits on  the DFSZ
axion  mass  derived from  white  dwarfs  using pulsating  stars  (top
section)  as well  as employing  the white  dwarf luminosity  function
(bottom section),  and including  the results  derived in  the present
paper.  The different axion-mass limits are illustrated
in Fig. \ref{figure6}. It is  worth noting  that the  measured
value  of the  rate of
period change of the 215~s  mode of G117$-$B15A has been substantially
changing over the  years due to the introduction  of corrections, like
the  proper motion  correction, and  the fact  that the  phase of  the
periodicity  is  affected  by  jumps caused  by  low-amplitude  modes,
contamination by  G117$-$B15B (the  companion star), use  of different
telescopes and apertures, and  several other subtle effects \cite{RN},
until  it  reached a  value  of  $\dot{\Pi}=  (4.19 \pm  0.73)  \times
10^{-15}$~s/s      according     to      the     last      measurement
\cite{2012ASPC..462..322K}.   In  Table  \ref{table3}, a  label  ``T''
(``NT'') in  parenthesis means  that the  mode for  which the  rate of
period change  has been  measured is  (is not) a  mode trapped  in the
outer envelope of  the star.  In particular, the mode  at $\sim 215$~s
($k= 2,\ \ell= 1$) of G117$-$B15A and  the mode at $\sim 212$ s ($k= 2,\
\ell= 1$) of R548 are trapped modes. We remind here that the bounds on
the    axion    mass    derived     using    these    trapped    modes
\cite{2012MNRAS.424.2792C,2012JCAP...12..010C} crucially depend on the
nature of  the mode.  Specifically, a trapped  mode has  a theoretical
rate of period change that is smaller than the values corresponding to
the adjacent,  non trapped  modes.  A small  value of  the theoretical
$\dot{\Pi}$      results      in      a     large      axion      mass
\cite{2012MNRAS.424.2792C}.       It       has       been       argued
\cite{2014JCAP...10..069M} that this is  the origin of the discrepancy
between  the value  of $m_{\rm  a}$ derived  from the  pulsating stars
G117$-$B15A  and R548  and the  limits inferred  from the  white dwarf
luminosity           function          \cite{WDLF,2009JPhCS.172a2005I,
2014JCAP...10..069M},  that discard  axion  masses  larger than  $\sim
10$~meV.  Interestingly,  for the  case of  the pulsating  white dwarf
L19$-$2 analyzed  in the present  paper, the modes with  periods $\sim
113$~s  and  $\sim 192$~s,  for  which  a  rate  of period  change  of
$\dot{\Pi}=  (3.0  \pm 0.6)  \times  10^{-15}$~s/s  has been  measured
\cite{2015ASPC..493..199S}, seem to be non-trapped modes, according to
our asteroseismic  model. However,  our analysis yet supports upper
limits  for the  axion mass  that  are substantially  higher than  the
largest bounds obtained  by using the white  dwarf luminosity function
($\sim 10$~meV). Note  that,  even if the uncertainties in the
theoretical  rates of  period change  were reduced  to zero,  we would
obtain upper limits for the mass of the axion that are still above the
maximum mass  allowed by studies  based on the luminosity  function of
white dwarfs. Nevertheless, it should be  kept in mind that the errors
in  the  observational determination  of  the  white dwarf
luminosity function are  poissonian. When  the  remaining  uncertainties
in  thedetermination of this function are taken into account this upper bound
considerably weakens.

In the  same vein, attention has been drawn  \cite{RN} to the
possibility of obtaining bounds to the axion mass from pulsating white
dwarfs with He-rich  atmospheres, the so-called DBV  class of variable
white dwarfs.   Recently, an upper limit  to the mass of  the axion of
$m_{\rm  a} \cos^2  \beta \sim 19.5$~meV ($g_{\rm ae}\sim 5.5 \times 10^{-13}$)
has been  obtained \cite{Battich}
using the DBV  star PG~1351+489 for which a value  of $\dot{\Pi}= (2.0
\pm     0.9)     \times     10^{-13}$~s/s     has     been     derived
\cite{2011MNRAS.415.1220R}. This upper limit seems  to be in line with
the predictions  of the analysis  based on the luminosity  function of
white dwarfs, with the previous ones derived employing G117$-$B15A and
R548 \cite{BKea08,2012MNRAS.424.2792C,  2012JCAP...12..010C}, and also
with those obtained in this paper for L19$-$2.

In  view  of  the  above  discussion,  and  taking  into  account  the
uncertainties still plaguing  both methods, it seems  that the results
obtained   from  the   luminosity   function  of   white  dwarfs   and
asteroseismic models of variable  of white dwarfs are concordant
at the $1\sigma$  level, although additional efforts  are necessary to
reduce the  discrepancies.  More measurements  of the rates  of period
change for other pulsating white  dwarfs, together with improvements
in the
asteroseismic models to reduce the uncertainties coming from the
previous evolution, will enable the determination of new and more
precise bounds on the axion mass.   At the same time, additional work
will be necessary
as well  to reduce  the discrepancies  between the  different observed
white dwarf luminosity  functions, along with better  estimates of the
statistical  uncertainties,   and  more  reliable   cooling  sequences
\cite{2014JCAP...10..069M}. All these  future improvements, will allow
us to better constrain the axion mass.


\section*{Acknowledgments}

 We wish to thank our anonymous referee for the 
constructive comments and suggestions that improved the 
original version of the paper. 
Part of  this work was  supported by  AGENCIA through the  Programa de
Modernización Tecnológica BID 1728/OC-AR,  by the PIP 112-200801-00940
grant  from  CONICET, by  MINECO  grant  AYA2014-59084-P, by  the
AGAUR, by CNPq and PRONEX-FAPERGS/CNPq (Brazil). This research
has made use of NASA Astrophysics Data System.

\bibliographystyle{JHEP}
\bibliography{L192}


\end{document}